# Dense Dilated UNet: Deep Learning for 3D Photoacoustic Tomography Image Reconstruction

Steven Guan, Ko-Tsung Hsu, Matthias Eyassu, and Parag V. Chitnis

*Abstract*—In photoacoustic tomography (PAT), the acoustic pressure waves produced by optical excitation are measured by an array of detectors and used to reconstruct an image. Sparse spatial sampling and limited-view detection are two common challenges faced in PAT. Reconstructing from incomplete data using standard methods results in severe streaking artifacts and blurring. We propose a modified convolutional neural network (CNN) architecture termed Dense Dilation UNet (DD-UNet) for correcting artifacts in 3D PAT. The DD-Net leverages the benefits of dense connectivity and dilated convolutions to improve CNN performance. We compare the proposed CNN in terms of image quality as measured by the multiscale structural similarity index metric to the Fully Dense UNet (FD-UNet). Results demonstrate that the DD-Net consistently outperforms the FD-UNet and is able to more reliably reconstruct smaller image features.

*Index Terms*—Image reconstruction, deep learning, tomography, photoacoustic imaging, biomedical imaging

## I. Introduction

PHOTOACOUSTIC imaging is an emerging and non-invasive hybrid imaging modality that combines the advantages of optical (e.g. high contrast and molecular specificity) and ultrasound (e.g. high penetration depth) imaging [1]. Given its unique use of light and sound, it has shown great potential for many preclinical and clinical imaging applications such as small animal whole-body imaging, breast and prostate cancer imaging, and image-guided surgery [2]–[5]. Moreover, multispectral photoacoustic imaging can be used for functional imaging such as measuring blood oxygen saturation and metabolism in biological tissues [6]. With the ability to provide both structural and functional information, photoacoustic imaging can reveal novel insights into biological processes and disease pathologies [7].

In photoacoustic tomography (PAT), a semi-transparent medium is illuminated by a short-pulsed laser. Chromophores (i.e., optically absorbing molecules) within the medium become excited and undergo thermoelastic expansion, which results in the generation of acoustic waves that can be detected using an array of acoustic sensors surrounding the medium [1]. The goal of PAT image reconstruction is to recover the initial pressure distribution (i.e., an image of chromophore distribution in the medium) from the measured time-dependent signal. This is a well-studied inverse problem that can be solved using analytical solutions, numerical methods, and model-based iterative methods [8]–[12].

In an ideal scenario, the sensor array is comprised of a large number of sensors and has a "full-view" of the imaging target, meaning the sensor array fully surrounds the medium. This would allow for the acquisition of complete data and the reconstruction of high-quality images using standard reconstruction methods. However, building an imaging system with these specifications is often prohibitively expensive, and in many *in vivo* applications, the sensor array can only partially surround the medium. These limitations result in the acquisition of incomplete data characterized by sparse spatial sampling and limited-view detection of the acoustic waves. Reconstructing images from sub-optimally acquired data results in strong streaking artifacts that severely degrades image quality and interpretability.

Advanced reconstruction methods such as iterative model-based methods can be used to reconstruct higher quality images with fewer artifacts from incomplete data [13], [14]. These methods use an explicit model of photoacoustic wave propagation and incorporate prior knowledge of the expected solution into the reconstruction process. However, iterative methods are computationally expensive due to the need for repeated evaluations of the forward and adjoint operators [15].

Given the success of deep learning in computer vision tasks such as classification and segmentation, there is increasing interest in applying these data-driven techniques for tomographic image reconstruction problems [16], [17]. By learning from data, deep learning techniques can efficiently overcome limitations in traditional image reconstruction techniques. Moreover, a trained deep learning model can reconstruct an image with relatively little computation and is well-suited for applications requiring near real-time imaging.

S. Guan is with the Bioengineering Department, George Mason University., Fairfax, VA 22031 USA. (e-mail: sguan2@gmu.edu) and The MITRE Corporation., McLean, VA, 22102 (e-mail: sguan@mitre.org). The author's affiliation with The MITRE Corporation is provided for identification purposes only and is not intended to convey or imply MITRE's concurrence with, or support for, the positions, opinions or viewpoints expressed by the author. Approved for Public Release; Distribution Unlimited. Case Number 21-0149. ©2021 The MITRE Corporation. All Rights Reserved.

K. Hsu, M. Eyassu, and P. Chitnis are with the Bioengineering Department, George Mason University, Fairfax, VA 22031 USA.

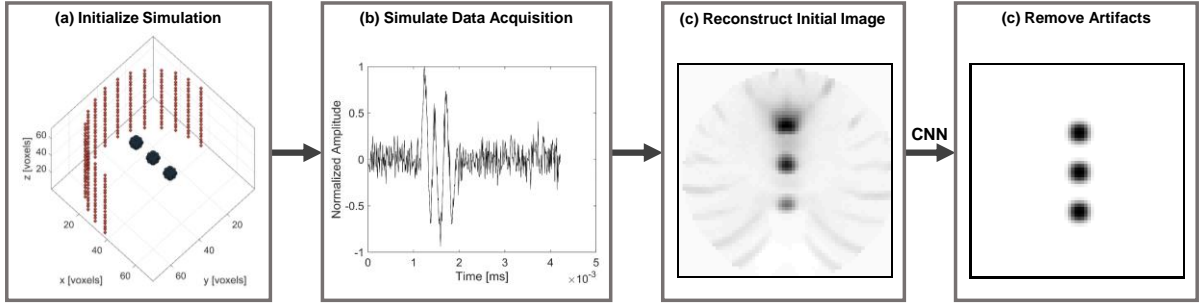

Fig. 1. Process diagram demonstrating the generation of sparse spatial sampling and limited-view 3D PAT data and Post-DL image reconstruction. (a) Simulation was initialized using a cylindrical sensor configuration (red elements) with a half-circle view and sparse spatial sampling to image spherical objects (black elements) in the center. (b) Example time-series data for a single sensor element with added Gaussian noise (25 dB PSNR). (c) Maximum intensity projection through the z-axis of the 3D image with artifacts when reconstructed using the time reversal method. (d) Maximum intensity projection through the z-axis of the 3D image without artifacts after post-processing using a CNN.

However, training a deep learning model requires large amounts of data which is often scarce or unavailable in biomedical imaging applications.

Many deep learning approaches have been developed for PAT image reconstruction [18]–[20]. Post-processing reconstruction (Post-DL) is the most widely used and has been previously demonstrated for removing artifacts and improving image quality in PAT and other imaging modalities such as CT and MRI [21]–[23]. In Post-DL, an initial image with artifacts is reconstructed from the time-series data, and a convolutional neural network (CNN) is applied as a post-processing step to remove artifacts [23], [24]. The main drawback of Post-DL is that potentially useful information in the time-series data is lost during the initial inversion. Other approaches (e.g. *Pixel-DL* and the *upgUNET)* improve upon Post-DL by replacing the initial inversion with a data pre-processing step to provide a more informative input for the CNN [25], [26]. Direct-learned reconstructions seeks to reconstruct an image directly from the time-series data with a CNN but often underperform compared to Post-DL [18]. Among the different approaches, model-based reconstruction was shown to outperform other deep learning approaches [27]. Like iterative reconstruction, this approach uses an explicit model of photoacoustic wave propagation, but the prior constraints are instead learned from data. The improved performance comes at the cost of increased computational complexity and slower image reconstruction.

In this work, the Post-DL approach is followed for 3D PAT reconstruction of sparse imaging targets in applications requiring fast image reconstruction. Although Pixel-DL has been shown to outperform Post-DL in 2D PAT, it is not suitable for 3D PAT imaging due to the large memory requirement and computational cost for manipulating the 4D pre-processed data array. We propose a modified CNN architecture termed Dense Dilated UNet (DD-Net) for 3D PAT imaging of sparse targets in a heterogeneous medium. This work builds upon the well-known UNet CNN architecture for biomedical imaging by incorporating dense connectivity and dilated convolutions throughout the network. Dense connectivity enables the CNN to learn more diverse feature sets by mitigating the need to relearn redundant features and enhancing information flow [28]. Dilated convolutions expand the CNN's effective receptive field without loss of resolution or coverage for learning multi-scale context [29].

## II. METHODS

### A. PAT Signal Generation and Image Reconstruction

The photoacoustic signal is generated by irradiating the medium with a nanosecond laser pulse. Chromophores in the tissue undergo thermoelastic expansion and generate acoustic pressure waves. Assuming negligible thermal diffusion and volume expansion during illumination, the initial acoustic pressure $x$ can be defined as

$$x(r) = \Gamma(r)A(r) \qquad (1)$$

where $A(r)$ is the spatial absorption function and $\Gamma(r)$ is the Grüneisen coefficient describing the conversion efficiency from heat to pressure [30]. The photoacoustic pressure wave $p(r, t)$ at position $r$ and time $t$ can be modeled as an initial value problem for the wave equation, in which $c$ is the speed of sound [31].

$$(\partial_{tt} - c_0^2\Delta)p(r, t) = 0, \quad p(r, t = 0) = x,$$
$$\partial_t p(r, t = 0) = 0 \quad (2)$$

Sensors located along a measurement surface, surrounding the medium, are used to measure a time-series signal (Fig. 1a). The linear operator $\mathcal{M}$ acts on $p(r, t)$ restricted to the boundary of the computational domain $\Omega$ over a finite time $T$ and provides a linear mapping from the initial pressure $x$ to the measured time-dependent signal $y$ (Fig. 1b).

$$y = \mathcal{M}_{p|\partial\Omega\times(0,T)} = Ax \quad (3)$$

Among the many reconstruction methods, time reversal is a robust method that works well for any arbitrary detection geometry and heterogenous mediums [10], [12], [32]. An image is formed by running a numerical model in which the measured time-series data is transmitted into the computational medium but in a time-reversed order. An artifact-free image can be reconstructed if the time-series data was measured using a sensor array with a high number of sensors and a full-view, and if the acoustic properties of the medium are known *a priori*. Reconstruction from incomplete data and the assumption of a homogeneous medium results in images with streaking artifacts and blurring.

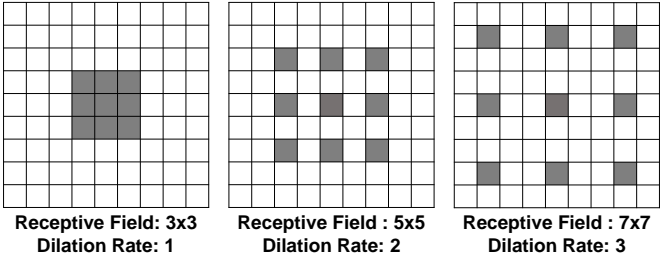

Fig. 2. In a dilated convolution, the effective receptive field of the convolution operation is enlarged by inserting gaps between the kernel weights of a 3x3 filter based on the dilation rate.

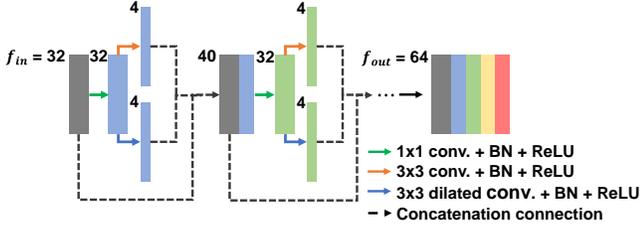

Fig. 3. Four layered dense dilation block with $k = 8$ and $f = 64$. In a dense dilation block, features learned from each convolutional layer are concatenated together with the input. Features are learned using both the standard and dilated convolution.

### B. Deep Learning Framework

An image containing streaking artifacts is initially reconstructed from the incomplete time-series data using time reversal (Fig. 1c). A CNN is then applied as a post-processing step to remove artifacts and improve image quality (Fig. 1d). This task can be formulated as a supervised learning problem, in which the CNN learns a function that maps the input, an image with artifacts, to the desired output, an artifact-free image [33]. The CNN is trained on paired examples of the initial time reversal reconstruction and the ground truth image.

### C. Dilated Convolutions

The dilated convolution, also known as the atrous convolution, is an extension of the standard convolution, in which the convolutional filter is upsampled by inserting zeros between the weights [34]. In the 1-D case of a dilated convolution, the output $o$ at location $i$ with a filter $w$ of size $S$ and dilation rate $r$ for an input $f$, can be represented as

$$o[i] = \sum_{s=1}^{S} f[i + r \cdot s] w[i] \tag{1}$$

When the dilation rate is one, the dilated convolution is equivalent to a standard convolution. A key advantage in using dilated convolutions is that the receptive field of the convolution operation can be enlarged without requiring additional training parameters (Fig.2). The receptive field describes the area of an image that can be viewed by an artificial neuron to extract information. A larger receptive field is needed to learn multi-scale features which is conventionally achieved by connecting successive convolutional layers in a cascade and using max pooling layers to spatially down sample the image [34]. Dilated convolutions allow the CNN to more efficiently learn multi-scale features without a rescaled image and loss of resolution. Cascaded dilated convolutions also expand the receptive field exponentially, whereas, cascaded standard convolutions expands it linearly [35].

However, it has been observed that the use of dilated convolutions results in "gridding artifacts" [34], [36]. Because of the zero-padded gaps in the convolutional filter, adjacent units in the output are calculated from completely separate inputs. Therefore, gridding artifacts occur when the image or feature map has higher-frequency content than the sampling rate of the dilated convolution [29]. Artifacts tend to be more severe for larger dilation rates and with cascaded dilated convolutions.

### D. Dense Dilation Blocks

For increasingly complex tasks, a deeper CNN with more convolutional layers is often needed to improve model performance. However, deeper networks suffer from the vanishing gradient problem, where the gradient is diminished as it is backpropagated through multiple layers [37], [38]. Trainable parameters in the earlier layers may fail to converge to optimal values resulting in suboptimal model performance. Dense connectivity addresses this problem by introducing numerous concatenation connections between convolutional layers which enable gradient information to directly flow into earlier layers [28].

In a dense block, the goal is to learn a total of $f$ features from the input features. This is achieved by iterating through several steps, where $k$ additional features are learned at each step. The key feature of dense connectivity is that earlier convolutional layers are connected to all subsequent layers by channel-wise concatenation [28], [39]. Each successive step learns additional features based on the original input provided and other features learned in previous layers. This removes the need to learn redundant features and promotes learning a more diverse set of features.

In this work, the dense block was modified to use both the standard and dilated convolutions (Fig. 3). At each step, $k/2$ features are learned with a standard convolution and the remaining features are learned using dilated convolutions with a dilation rate, $r$. This combination was used to mitigate potential gridding artifacts that may arise from using solely dilated convolutions. Furthermore, dilated convolutions, having a larger receptive field, can efficiently learn global context. Whereas, standard convolutions, having a denser receptive field, can efficiently learn local context.

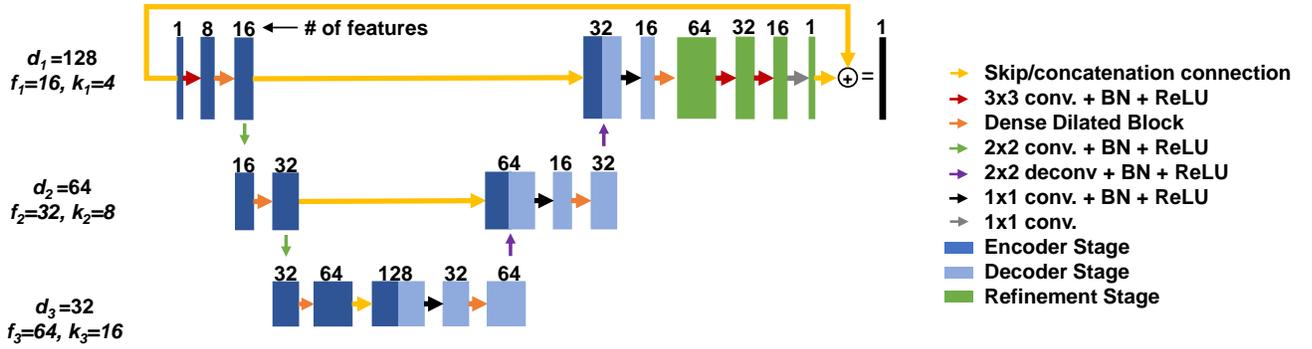

Fig. 4. Proposed DD-UNet architecture that incorporates dense connectivity and dilated convolutions throughout the UNet. In addition to the encoder and decoder structure of the standard UNet, several convolutional layers collectively termed the "refinement stage" were included following the decoder stage. Hyperparameters for the illustrated architecture are $k_1 = 8$ and $f_1 = 16$ for an input image of size 128x128x128 pixels.

## E. Dense Dilated UNet

The DD-Net is an enhanced version of our previous work, the Fully Dense U-Net (FD-UNet), which was shown to be superior to the standard UNet. The DD-Net follows an "encoder-decoder-refinement" structure [40]. The key innovation in the DD-Net is the unique use of dense dilation blocks to leverage the benefits of dense connectivity and dilated convolutions (Fig. 4). Max pooling layers were removed in the DD-Net because they often result in high frequency content that may cause gridding artifacts [29]. They were replaced by 2x2 convolutional layers with a stride of two which allow the CNN to learn a more useful transformation for spatial down sampling. A shallow "refinement" network comprised of a dense dilation block and two 3x3 convolutional layers was also added to the end of the original network. These additional layers allow the CNN to further refine the image and remove artifacts at the highest spatial resolution. In the original FD-UNet, only a 1x1 convolutional layer was applied at the end of the decoding stage to form the final image. Addition of a "refinement" stage has been shown to improve model performance [41].

## F. Generating Training and Testing Data

Synthetic sphere phantoms were generated by placing 25-50 spheres with randomly selected center coordinates, radius (range 5 to 10 pixels), and magnitude (range 1 to 5) in a 128x128x128 pixels image. Resulting images were smoothed with a 5x5 moving average filter. This process was repeated to create a training dataset with 1000 images and a testing dataset with 500 images.

The "ELCAP Public Lung Image Database" is comprised of 50 whole-lung CT scans that were obtained within a single breath hold [42]. These scans were split into training (N=40) and testing groups (N=10) and were used to generate additional training and testing data via data augmentation. Each 3D scan had dimensions of 512x512x288 pixels. First, the lungs were segmented from the CT scan using active contours with the Chan-Vese algorithm [43]. Next, the Frangi vesselness filter was applied to suppress background noise and segment vessel-like structures in the lungs [44]. 3D vasculature phantoms were then procedurally generated by randomly rotating the filtered 3D images along each axis and then sampling a 128x128x128 pixels image. This data augmentation process was repeated to create a training dataset with 1000 images and a testing dataset with 500 images.

Synthetic breast vasculature phantoms were created using an analytic approach to generate random but realistic anatomical structures within a predefined breast volume [45]. This method was originally developed for the "Simulated Virtual Imaging Clinical Trial for Regulatory Evaluation" project which sought to demonstrate *in silico* imaging trials and imaging computer simulation tools as a viable source of evidence for the regulatory evaluation of imaging devices. From this approach, 400 different breast phantoms with dimensions of 718x796x506 were generated. These phantoms were split into training (N=300) and testing groups (N=100) and were used to generate additional training and testing data via a similar data augmentation strategy as described earlier. A training dataset with 1000 images and a testing dataset with 500 images of breast vasculature were created.

The MATLAB toolbox k-WAVE was used to simulate photoacoustic data acquisition using an array of acoustic sensors arranged in a cylindrical geometry [46]. The sensor array is essentially a linear array with 128 elements along the z-axis that is repeated at equally spaced intervals along a half-circle in the x-y plane (Fig. 1a). In order to have experiments with varying levels of sparsity, three different sensor arrays with sampling at 10, 20, 30 angles in the x-y plane were used for simulations. Having fewer angles sampled results in more severe sparse spatial sampling artifacts.

Training and testing phantoms were normalized (values between 0 and 1) and treated as a photoacoustic source distribution on a computation grid of 128x128x128 pixels. The medium was assumed to be non-absorbing and heterogeneous, in which the background had a speed of sound of 1480 m/s and density of 1000 kg/m$^3$ while the vasculature had a speed of sound of 1570 m/s and density of 1060 kg/m$^3$. The time reversal method in the k-WAVE toolbox was used for reconstructing an initial image from the simulated photoacoustic time series data. In the scenario of *in vivo* imaging, the spatial distribution of the speed of sound and density is unknown. Thus, the reconstruction was completed assuming a homogeneous

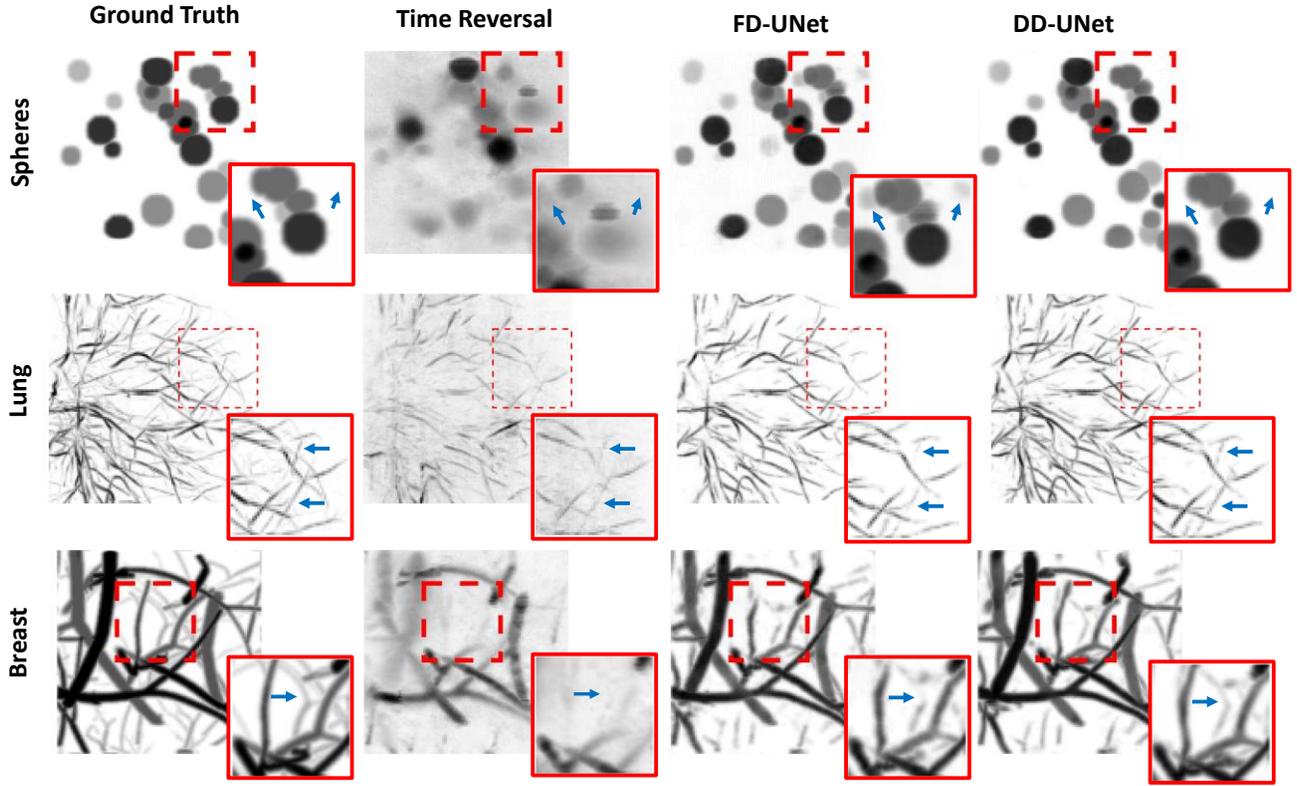

Fig. 5. Example ground truth and reconstructed images using the time reversal, FD-UNet, and the DD-UNet (dilation rate = 2) methods for three different imaging phantoms reconstructed with a sampling sparsity of 30 angles. The smaller image with a solid red border is an enlarged sub-image from the region designated by the dashed red line. The blue arrows highlight key differences between the reconstructed images (Top) Spheres phantom. The FD-UNet image incorrectly had spheres in the background that were not in the ground truth or DD-UNet images. (Middle) Lung vasculature phantom. The small vessels were more visible and clearer in the DD-UNet image than the FD-UNet image. (Bottom) Breast vasculature phantom. A small vessel that was not in the time reversal image was recovered in the CNN images but appeared to be sharper in the DD-UNet image.

medium with a speed of sound of 1480 m/s and density of 1000 kg/m³.

### G. Evaluating Image Quality for Sparse Images

To evaluate image reconstruction quality, the multi-scale structural similarity index metric (MS-SSIM) was used to compare the reconstructed image to the ground truth image [47]. MS-SSIM is a composite metric that measures similarities between two images in terms of contrast, luminance, and structure at multiple spatial scales. Similarities are calculated based on a local neighborhood of pixels, and a global value is reported by averaging the neighborhood values. MS-SSIM is superior to other metrics such as the standard SSIM and peak-signal-to-noise-ratio (PSNR) for evaluating image quality in 3D images with sparse imaging targets that occupy a small fraction of the space in the medium. The main drawback of the SSIM and PSNR metrics is that image quality is only evaluated at a single spatial scale. Therefore, these metrics are heavily biased by how well artifacts were removed from the background and only weakly associated with how well sparse structures were reconstructed.

### H. Deep Learning Implementation

The CNNs are implemented in Python 3.7 with TensorFlow v2.1, an open source library for deep learning [48]. Training and evaluation of the network is performed on an NVIDIA V100 GPU. The CNNs were trained using the Adam optimizer to minimize the mean squared error loss with an initial learning rate of 1e-4 and a batch size of two images for 500 epochs. The same hyperparameters (i.e., f=16, k=4, and L=3) were used for both CNNs, and the DD-UNet had a dilation rate of two. Training each CNN required approximately one day to complete. A separate CNN was trained for each CNN architecture, PAT imaging system configuration, and dataset. The FD-UNet (120,000) and DD-UNet (150,000) had a similar number of parameters.

### III. RESULTS

In silico experiments were performed using three different sparse imaging phantoms (i.e., spheres, lung vasculature, and breast vasculature) to evaluate the FD-UNet and DD-UNet for Post-DL image reconstruction. Given an initial time reversal reconstructed image, the CNNs were tasked with removing artifacts arising from sparse spatial sampling, limited-view detection, and an unknown heterogeneous medium. The MS-SSIM metric was used to evaluate image quality by comparing the reconstructed images to the ground truth image for N=500 testing image pairs in each dataset. Using Otsu's method for automated thresholding and binarization, the imaging targets were estimated to on average occupy 3-5% of the space in the imaging phantoms.

### A. Visual Comparison of CNN Images

In this initial experiment, the imaging system used a half-view cylindrical sensor array to sample the acoustic waves at 30 equally spaced angles. In general, it is difficult to visually identify differences between the FD-UNet and DD-UNet image reconstructions because the differences are subtle. Maximum intensity projections are convenient for visualizing 3D features in a 2D image, but only differences between the most prominent features can be seen in the projections. In the representative examples, both CNNs produce images that are of higher quality than the time reversal reconstructed images (Fig. 5). Spheres and vessels not visible in the time reversal reconstructions can be clearly seen in the CNN reconstructions. While both CNNs remove most artifacts and accurately reconstruct the larger and more prominent image features. The DD-UNet was observed to be better in reconstructing the smaller image features. These features are typically either missing or inaccurately reconstructed in the FD-UNet images. Furthermore, the FD-UNet occasionally mistakenly interpreted artifacts observed in the time reversal reconstruction as a true imaging target.

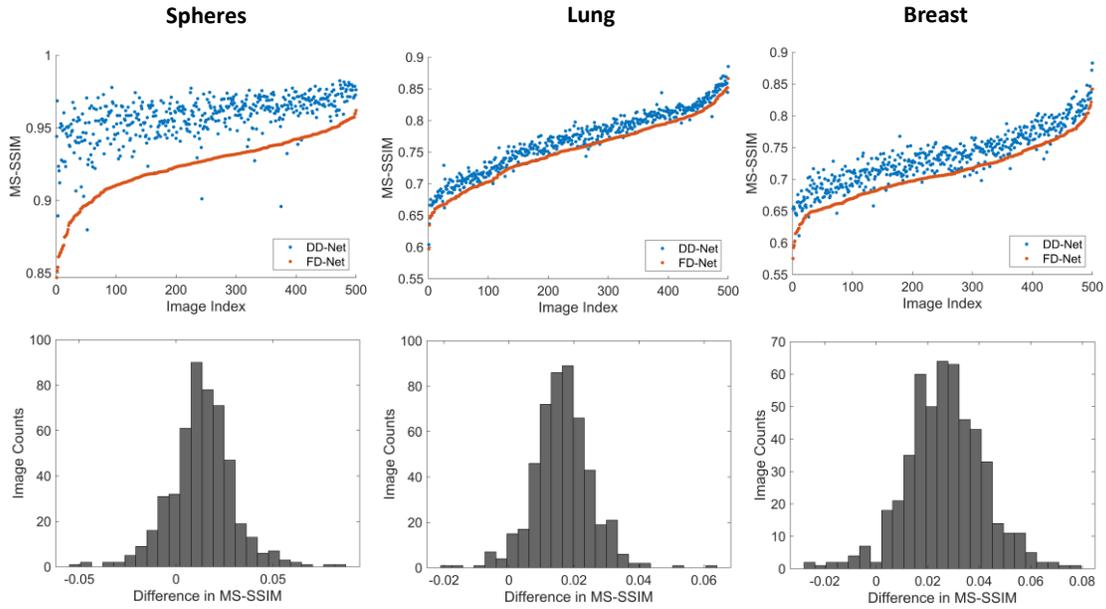

Fig. 6. (Top) Scatter plots for comparing the MS-SSIM of the FD-UNet and DD-UNet (dilation rate = 2) image reconstructions for each imaging phantom. For improved visualization, the image index was defined based on the sorted order of the MS-SSIM scores for the FD-UNet. (Bottom) Histogram showing the difference in MS-SSIM for the same test image between the DD-UNet and FD-UNet for each imaging phantom. A positive difference indicates that the DD-UNet reconstructed a higher quality image. Results shown are for a sparsity level of 30 angles sampled.

TABLE I
Image Reconstruction Quality for Different Levels of Sparsity

| Image Phantom | Angles Sampled | MS-SSIM[a] | | | Difference in MS-SSIM[b] | | |
|---|---|---|---|---|---|---|---|
| | | Time Reversal | FD-UNet | DD-UNet | FD-UNet and Time Reversal | DD-UNet and Time Reversal | DD-UNet and FD-UNet |
| Spheres | 10 | 0.073 ± 0.015 | 0.824 ± 0.049 | 0.837 ± 0.051 | 0.751 ± 0.056 | 0.764 ± 0.057 | 0.013 ± 0.029 |
| | 20 | 0.181 ± 0.026 | 0.918 ± 0.025 | 0.931 ± 0.024 | 0.737 ± 0.038 | 0.750 ± 0.038 | 0.013 ± 0.017 |
| | 30 | 0.297 ± 0.028 | 0.925 ± 0.021 | 0.958 ± 0.014 | 0.627 ± 0.034 | 0.661 ± 0.032 | 0.033 ± 0.017 |
| Lung | 10 | 0.234 ± 0.077 | 0.558 ± 0.051 | 0.568 ± 0.055 | 0.324 ± 0.095 | 0.334 ± 0.098 | 0.010 ± 0.013 |
| | 20 | 0.320 ± 0.083 | 0.665 ± 0.054 | 0.707 ± 0.050 | 0.344 ± 0.104 | 0.387 ± 0.102 | 0.043 ± 0.014 |
| | 30 | 0.384 ± 0.087 | 0.753 ± 0.049 | 0.769 ± 0.049 | 0.368 ± 0.104 | 0.385 ± 0.105 | 0.017 ± 0.009 |
| Breast | 10 | 0.101 ± 0.024 | 0.469 ± 0.055 | 0.500 ± 0.059 | 0.368 ± 0.071 | 0.399 ± 0.075 | 0.031 ± 0.015 |
| | 20 | 0.157 ± 0.024 | 0.659 ± 0.050 | 0.690 ± 0.047 | 0.502 ± 0.065 | 0.533 ± 0.061 | 0.031 ± 0.016 |
| | 30 | 0.243 ± 0.027 | 0.710 ± 0.044 | 0.736 ± 0.044 | 0.466 ± 0.056 | 0.493 ± 0.056 | 0.027 ± 0.015 |

[a] Mean and standard deviation of MS-SSIM scores of images reconstructed for each dataset.  [b] Mean and standard deviation of the differences in MS-SSIM between pairs of image reconstructions for each dataset. The DD-UNet had a dilation rate of two.

### B. Quantitative Comparison of CNN Performance

Images reconstructed with the CNNs had MS-SSIM scores ranging from 0.85 to 0.97 (spheres), 0.59 to 0.88 (lung), and 0.57 to 0.88 (breast). More complex imaging phantoms (e.g. containing more spheres or vessels) or those with imaging targets further away from the imaging sensor array typically resulted in reconstructions with lower scores. For all experiments performed, the DD-UNet consistently outperformed the FD-UNet when comparing the MS-SSIM for the same image reconstructed (Fig. 6). There were only a few instances in which the FD-UNet reconstructed the test image with a higher MS-SSIM than DD-UNet. Interestingly, the degree of improvement in MS-SSIM by the DD-UNet appeared to have a stochastic nature. When examining the distribution of differences in MS-SSIM between the CNNs, the DD-UNet outperformed the FD-UNet with a mean and standard deviation of 0.033 ± 0.016 (spheres), 0.017 ± 0.009 (lung), and 0.027 ± 0.015 (breast) (Fig. 6).

## C. CNN Performance at Different Levels of Sparsity

By decreasing the number of angles sampled, the acoustic waves were more sparsely sampled. This resulted in increasingly severe streaking artifacts and thus in a more difficult problem for the CNNs to overcome. As expected, the average MS-SSIM scores decreased as the number of angles sampled decreased for all reconstruction methods (Table I). Both CNNs consistently improved the time reversal reconstruction for all imaging phantoms and levels of sparsity tested. The large difference in MS-SSIM between the CNNs and time reversal reconstructions can be mostly explained by the fact that both CNNs were highly proficient at removing background artifacts and properly reconstructing the larger image features (Fig. 4). The DD-UNet was shown to significantly outperform the FD-UNet for all imaging phantoms and levels of sparsity tested (Wilcoxon matched-pairs signed rank test, p<0.01).

## IV. Discussion and Conclusion

In this work, we propose a modified CNN architecture termed DD-Net for 3D sparse and limited-view PAT image reconstruction that leverages the benefits of both dense connectivity and dilated convolutions through the unique use of dense dilation blocks. *In silico* experiments were performed with three different sparse phantoms (i.e., spheres, lung vasculature, and breast vasculature), and the DD-Net was demonstrated to be a superior CNN architecture compared to the FD-UNet. For all experiments performed, the DD-Net consistently reconstructed the image with a higher MS-SSIM by 0.01 to 0.03 depending on the phantom and level of sampling sparsity (Table I). Images reconstructed by the DD-UNet and FD-UNet did not have many large and obvious visual differences. However, the DD-UNet was observed to be able to reconstruct the smaller structures and finer details more accurately. For example, small vessels in the breast and lung vasculature phantoms that were missing or inaccurately reconstructed in the FD-UNet image could be seen more clearly in the DD-Net image (Fig. 4). These improvements were likely due to the expanded receptive field enabling the CNN to use more context in the image to reconstruct these finer features and the addition of a shallow network termed the refinement stage to further correct artifacts at the highest image resolution.

Choice of dilation rate for the DD-UNet depends on the imaging targets to be reconstructed and size of the imaging volume. For example, larger volumes with predominantly bigger image features may benefit from a larger dilation rate since more global context is available to the CNN. However, increasing the dilation rate does not necessarily lead to improved performance since gridding artifacts can become more severe. Some image features might also be smaller than the zero-filled gaps in a large receptive field leading to a loss of local context, but this issue is mitigated to a degree in the DD-UNet by using a combination of standard and dilated convolutions.

A limitation in applying deep learning for 3D PAT image reconstruction is the limited GPU memory. This forces the use of shallower CNNs with fewer convolutional layers, which constrains the representational power or complexity of the CNN and results in suboptimal model performance. For example, the CNNs in this work for 3D PAT have about ~$10^5$ parameters due to memory limitations, while an equivalent CNN for 2D PAT had ~$10^6$ or more parameters. Reconstructing larger volumes requires more memory and further limit the complexity of the CNN. Further work in developing more memory efficient CNN architecture or strategies for 3D PAT image reconstruction is needed to address this issue.

A key challenge in applying deep learning for *in vivo* PAT image reconstruction is the need for a large training dataset. Arbitrarily large synthetic training data can be generated using numerical phantoms and anatomical templates with data augmentation as demonstrated in this work. The synthetic data does need to properly capture the expected variations in artifacts observed in the experimental data. Therefore, the PAT simulation parameters need to be well-matched with the experimental conditions. Depending on the PAT imaging system, this can be a non-trivial task since it requires the system to be well-characterized.

The proposed Post-DL approach using the DD-Net can be used as a computationally efficient method for improving PAT image quality under limited-view and sparse sampling conditions. It can be applied to a wide variety of PAT imaging applications and allows for the development of more efficient data acquisition using fewer sensors without sacrificing image quality. This approach enables real-time PAT image rendering which would provide valuable feedback while imaging. The DD-Net can also be readily applied to image reconstruction problems in other imaging modalities (e.g., ultrasound and CT) and other biomedical imaging applications such as segmentation.

## V. Acknowledgements

This project was supported by resources provided by the Office of Research Computing at George Mason University (URL: https://orc.gmu.edu) and funded in part by grants from the National Science Foundation (Awards Number 1625039 and 2018631).